\documentclass[
showpacs,showkeys,preprintnumbers,aps,prd]{revtex4}

\usepackage{amsmath}
\usepackage{amssymb}
\usepackage{revsymb}
\usepackage{graphicx}
\usepackage{dcolumn}

\begin{document}

\title{Flavor analysis of nucleon electromagnetic form factors}
\author{
M. Rohrmoser$^1$, K.-S. Choi$^2$, and W. Plessas$^1$}

\address{$^1$ Theoretische Physik, Institut f\"ur Physik,
Karl-Franzens-Universit\"at Graz, Universit\"atsplatz 5, A-8010 Graz, Austria}
\address{$^2$ Department of Physics, Soongsil University, Seoul 156-743, Republic of
Korea}

\begin{abstract}
We have performed an evaluation of the individual up- and down-flavor contributions
to the nucleon electromagnetic form factors within a relativistic constituent-quark
model. It is found that the theoretical parameter-free predictions agree surprisingly
well with recent phenomenological data from a flavor decomposition of the proton and neutron
electromagnetic form factors for momentum transfers up to $Q^2\sim\;$3.5 GeV$^2$.
It means that in this regime three-quark valence degrees of freedom dominate
and other contributions, such as explicit mesonic effects, can at most play a minor role. 
\end{abstract}


\maketitle
By the advent of more and more precise experimental data on nucleon form factors,
especially from Jefferson Lab, it has recently become possible to reliably identify the various
$u$- and $d$-flavor contributions to the elastic proton and neutron electromagnetic form
factors $G_E^p$, $G_M^p$, $G_E^n$, and $G_M^n$~\cite{Cates:2011pz,Qattan:2012zf,Diehl:2013xca}.
The corresponding results cover the range of momentum transfers up to $Q^2 \sim 3.5$ GeV$^2$. Some essential observations thus gained have been described, e.g., in ref.~\cite{Cates:2011pz}
as:
\begin{itemize}
\item The $Q^2$ dependences of the ratios of the Pauli to Dirac form factors
      $F^q_2/F^q_1$ for the quark-flavor contributions $q$=$u,d$ to the nucleon form factors
      are practically constant for momentum transfers beyond $Q^2\sim\;$1.5 GeV$^2$, in
      contrast to the observed behaviors of the same ratios for both the proton and the
      neutron.
\item The $d$-flavor contributions $F^d_1$ and $F^d_2$ of the Dirac and Pauli form factors
      show a scaling behavior with $Q^4$ starting at momentum transfers of
      $Q^2\sim\;$1.5 GeV$^2$. However, this seems not to be confirmed by a very recent
      analysis in ref.~\cite{Qattan:2012zf}.
\item On the other hand, the $u$-flavor contributions $F^u_1$ and $F^u_2$ of the Dirac and
      Pauli form factors do not scale with $Q^4$ or drop off in the considered range of
      momentum transfers.         
\end{itemize}
In refs.~\cite{Cates:2011pz,Qattan:2012zf} one has also made comparisons to selective
theoretical results,
with no one being able to explain all the particular characteristics of the available
phenomenological data. Also some speculations are offered for the interpretation of the
new phenomenological insights, like a possible role/interplay of three- and five-quark
components in the nucleon states, quark-diquark clustering, or the onset of perturbative
quantum chromodynamics.

We have analyzed the various flavor ingredients in the nucleon electromagnetic form
factors within the relativistic constituent-quark model (RCQM) whose hyperfine interaction
is derived from Goldstone-boson exchange~\cite{Glozman:1998ag}. This type of dynamics stems
from the spontaneous breaking of chiral symmetry (SB$\chi$S) in low-energy quantum
chromodynamics (QCD) and produces a flavor-dependent spin-spin interaction that allows
to describe the baryon spectra in agreement with phenomenology, notably also with the
right level orderings in the nucleon and other baryon excitation spectra~\cite{Glozman:1998fs}.

The GBE RCQM has also been applied to produce covariant results for the nucleon
electromagnetic form factors~\cite{Wagenbrunn:2000es,Boffi:2001zb}, including the
electric radii and magnetic moments~\cite{Berger:2004yi}. The direct predictions
of the GBE RCQM have in all instances been found in remarkably good agreement
with existing experimental
data without the need of introducing any further parameterizations, such as
constituent-quark form factors or the like. These studies have
been performed in the framework of Poincar\'e-invariant quantum mechanics using its
point form~\cite{Dirac:1949cp}. The approach allows to calculate covariant observables,
since the Lorentz transformations (including rotations) can be constructed from
interaction-independent generators. This is even true for the electromagnetic current
assumed according to a spectator model~\cite{Melde:2004qu}. It has been shown that the
relativistic results obtained in this manner do not only fulfill all requirements of
the Poincar\'e algebra but also other necessary constraints, notably current
conservation~\cite{Melde:2007zz}. The GBE RCQM has been similarly successful
in predicting the axial and induced pseudoscalar form factors of the
nucleons~\cite{Boffi:2001zb,Glozman:2001zc}. Most recently one has obtained also the
axial charges of the other baryon ground states and their resonances and found them
in reasonable agreement with existing lattice QCD results~\cite{Choi:2010ty,Choi:2009pw}.
Similarly, the GBE RCQM has been able to provide a microscopic description of the $\pi NN$ as
well as $\pi N\Delta$ interaction vertices in accordance with their behavior expected from
phenomenology~\cite{Melde:2008dg}.

In view of the recently published phenomenological data, we have put
the GBE RCQM to the test of producing the flavor contributions to the nucleon
electromagnetic form factors. This is particularly interesting, as there has not yet
been any consistent theoretical explanation of the whole set of experimental data.
The calculations have been performed in complete accordance with the work in
refs.~\cite{Wagenbrunn:2000es,Boffi:2001zb,Berger:2004yi}. There one can also find detailed
descriptions of the formalism and the calculations of the nucleon elastic electromagnetic
form factors, which cannot be repeated here.

\begin{figure*}[t]
\centering
\includegraphics[clip=,width=9cm]{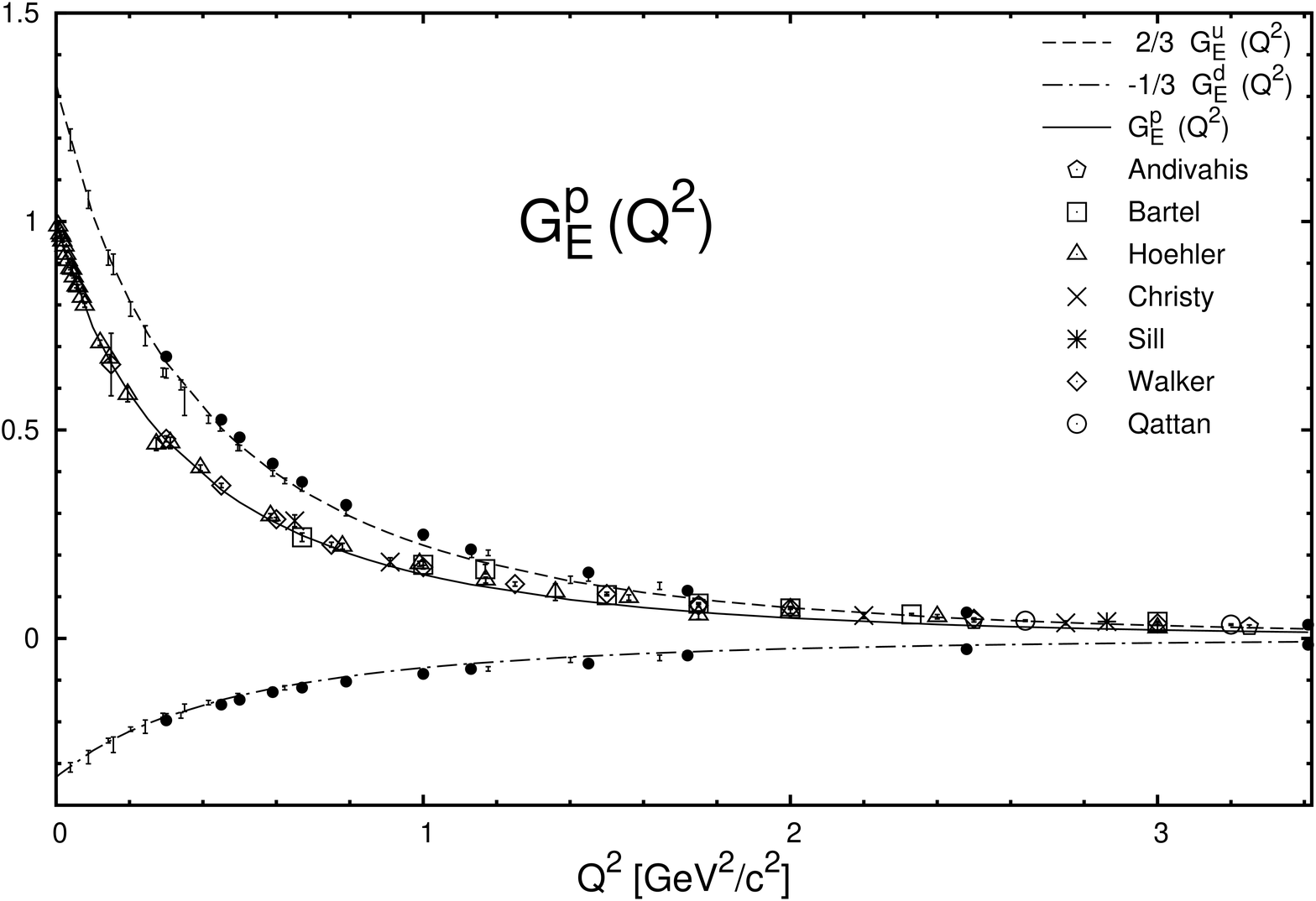} \hspace{-4mm}
\includegraphics[clip=,width=9cm]{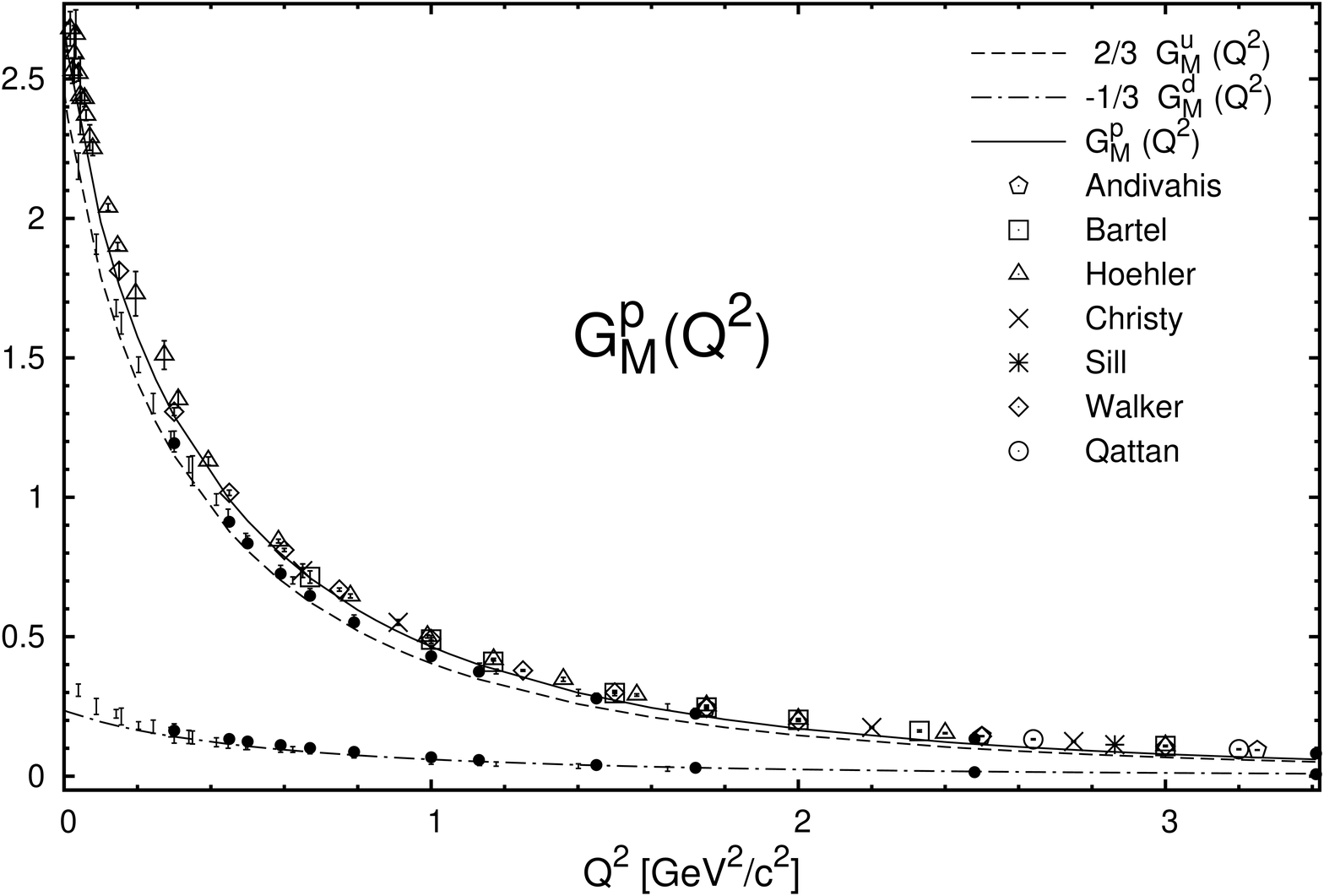}\\
\includegraphics[clip=,width=9cm]{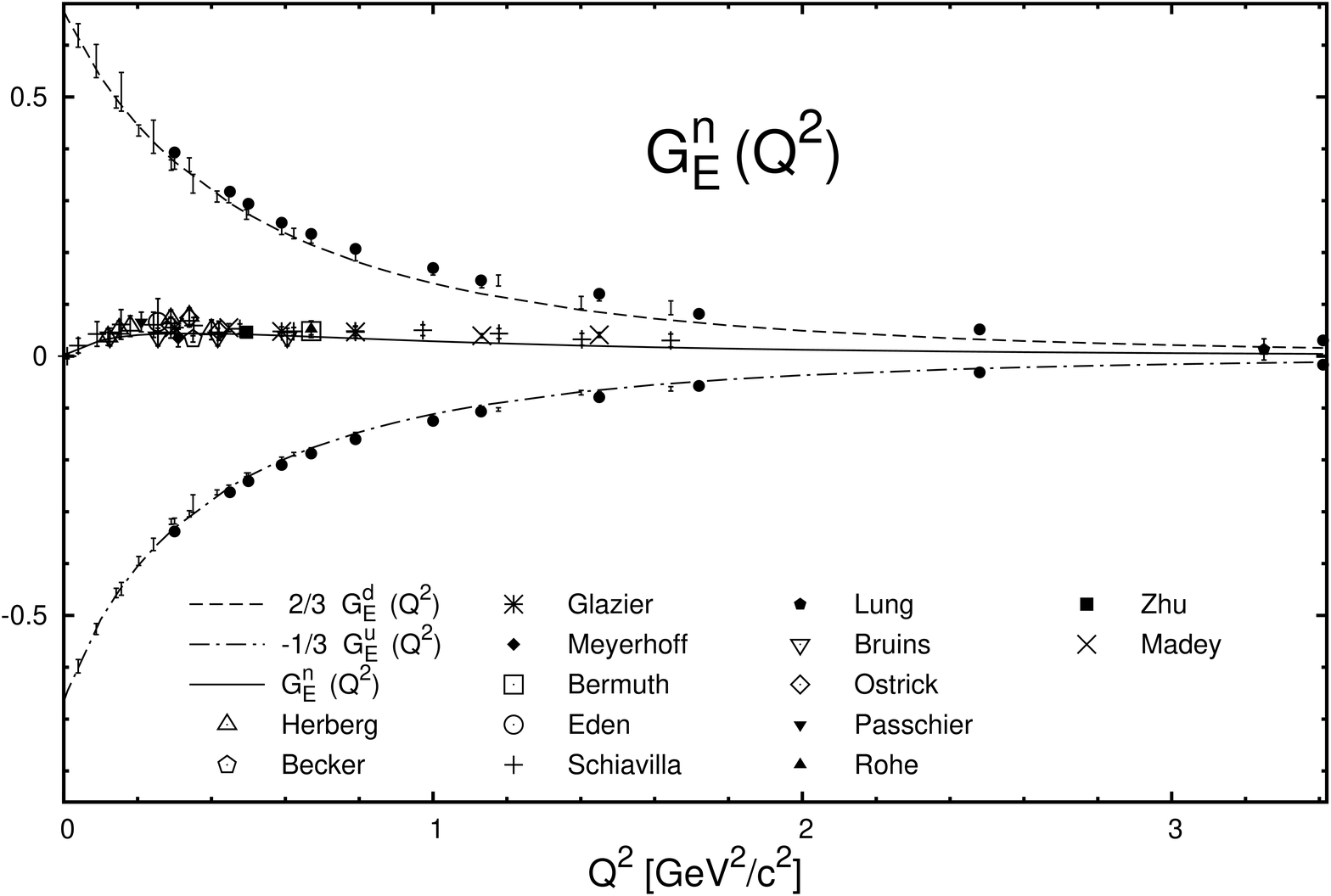}\hspace{-4mm}
\includegraphics[clip=,width=9cm]{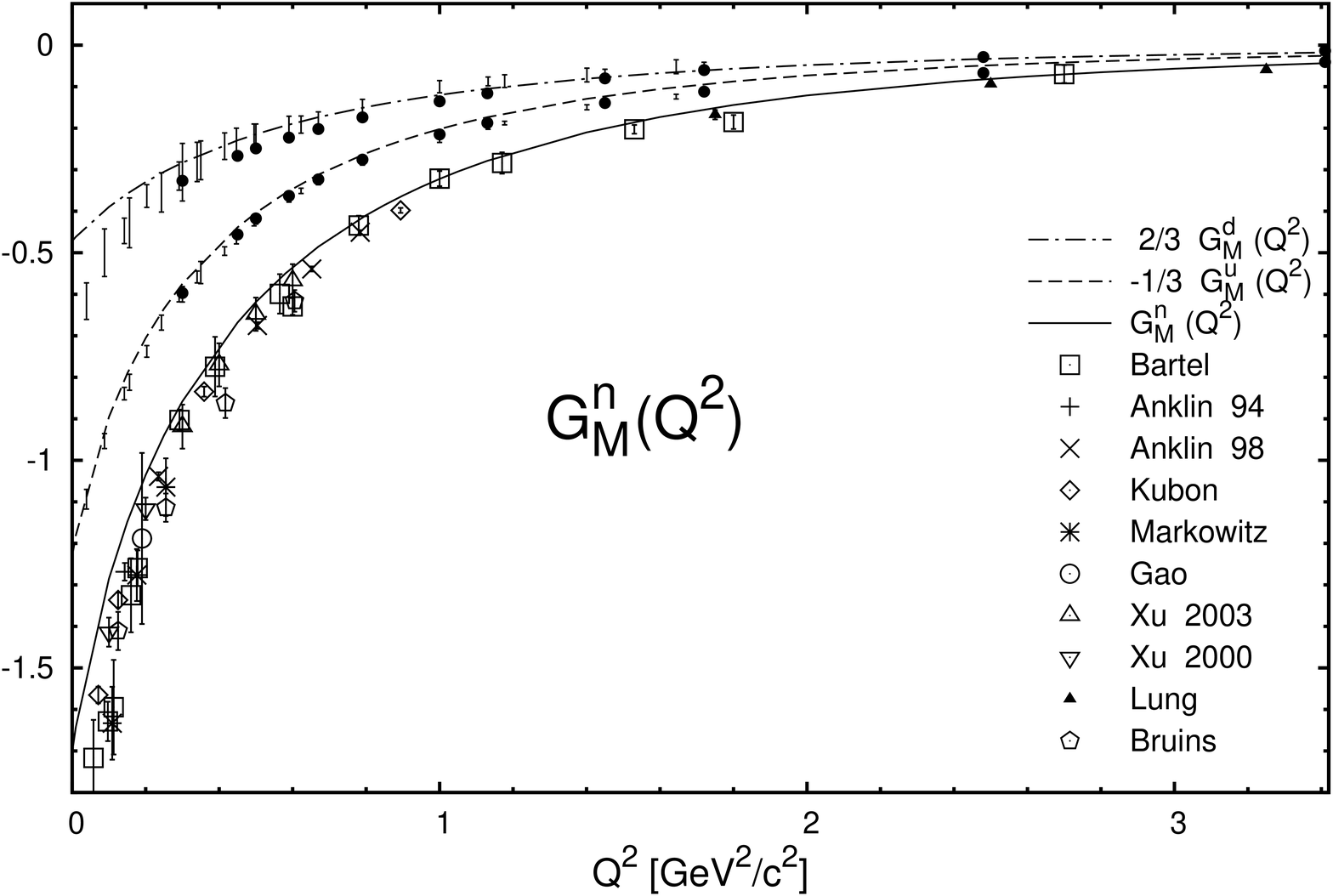}
\caption{$u$- and $d$-flavor contributions to the proton (upper panels) and neutron (lower
panels) electric and magnetic form factors as predicted by the GBE RCQM in comparison to
phenomenological data from refs.~\cite{Cates:2011pz} (filled circles) and~\cite{Diehl:2013xca}
(vertical bars) as well as global experimental data as indicated. The uncertainties in the data of ref.~\cite{Cates:2011pz} are practically not visible, as they do not exceed the
sizes of the circles; the uncertainties in the data of ref.~\cite{Diehl:2013xca} correspond
to the heights of the vertical bars. Notice that in case of the neutron the $u$- and
$d$-flavor contributions are here represented by $G_{E,M}^{d}$ and $G_{E,M}^{u}$, respectively;
cf. Eqs.~(\ref{u}) and~(\ref{d}).}
\label{ffs}
\vspace{-3mm}
\end{figure*}

Under charge symmetry the Sachs form factors of the nucleons are constituted as
\begin{eqnarray}
\label{electric}
G_E^p&=&\frac{2}{3}G_E^u-\frac{1}{3}G_E^d\,, \quad\quad
G_E^n=\frac{2}{3}G_E^d-\frac{1}{3}G_E^u\,, \\
\label{magnetic}
G_M^p&=&\frac{2}{3}G_M^u-\frac{1}{3}G_M^d\,, \quad \,\,
G_M^n=\frac{2}{3}G_M^d-\frac{1}{3}G_M^u \,,
\end{eqnarray}
where $G_{E,M}^u$ and $G_{E,M}^d$ are related to the $u$- and
$d$-flavor contributions in the proton and neutron form factors by:
\begin{eqnarray}
\label{u}
G_{E,M}^u(Q^2)&=&G_{E,M}^{u\,\,p}(Q^2)=2G_{E,M}^{d\,\,n}(Q^2)\,,\\
\label{d}
G_{E,M}^{d}(Q^2)&=&G_{E,M}^{d\,\,p}(Q^2)=\frac{1}{2}G_{E,M}^{u\,\,n}(Q^2)\,.
\end{eqnarray}

In Fig.~\ref{ffs} we show the results for the proton and neutron
electric and magnetic form factors as they are composed of the $u$- and $d$-flavor
contributions defined in Eqs.~(\ref{electric}) to (\ref{d}). Starting out
from zero momentum transfer, where the electric radii and magnetic moments are
reasonably well described~\cite{Berger:2004yi}, it is nicely seen, how the right
$Q^2$ dependence develops. Slight differences to the phenomenological data are only
visible for the anyway very small $u$-flavor contribution $\frac{2}{3}G_M^d$ to the
magnetic form factor of the neutron.

\begin{figure*}[t]
\includegraphics[clip=,width=9cm]{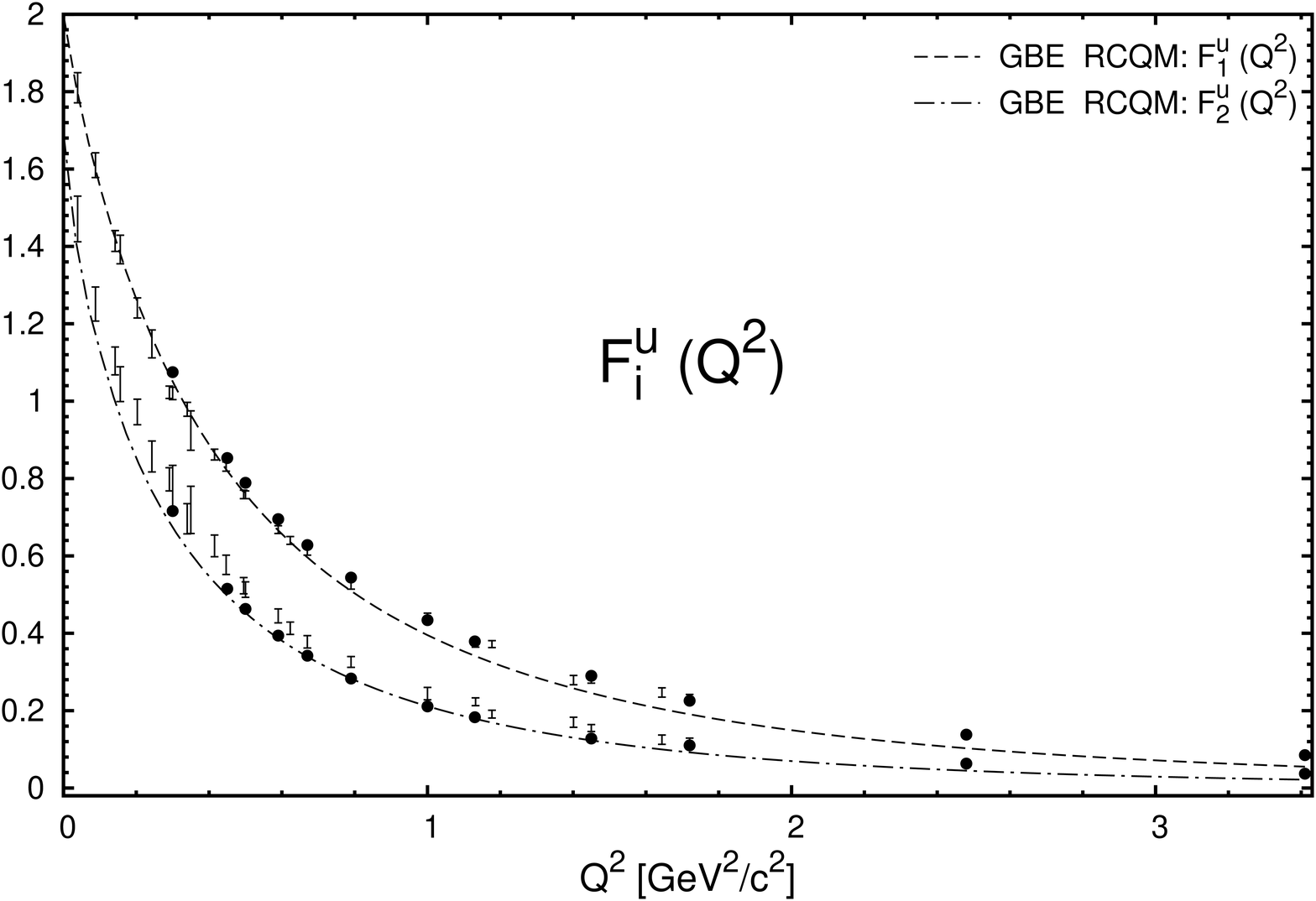} \hspace{-4mm}
\includegraphics[clip=,width=9cm]{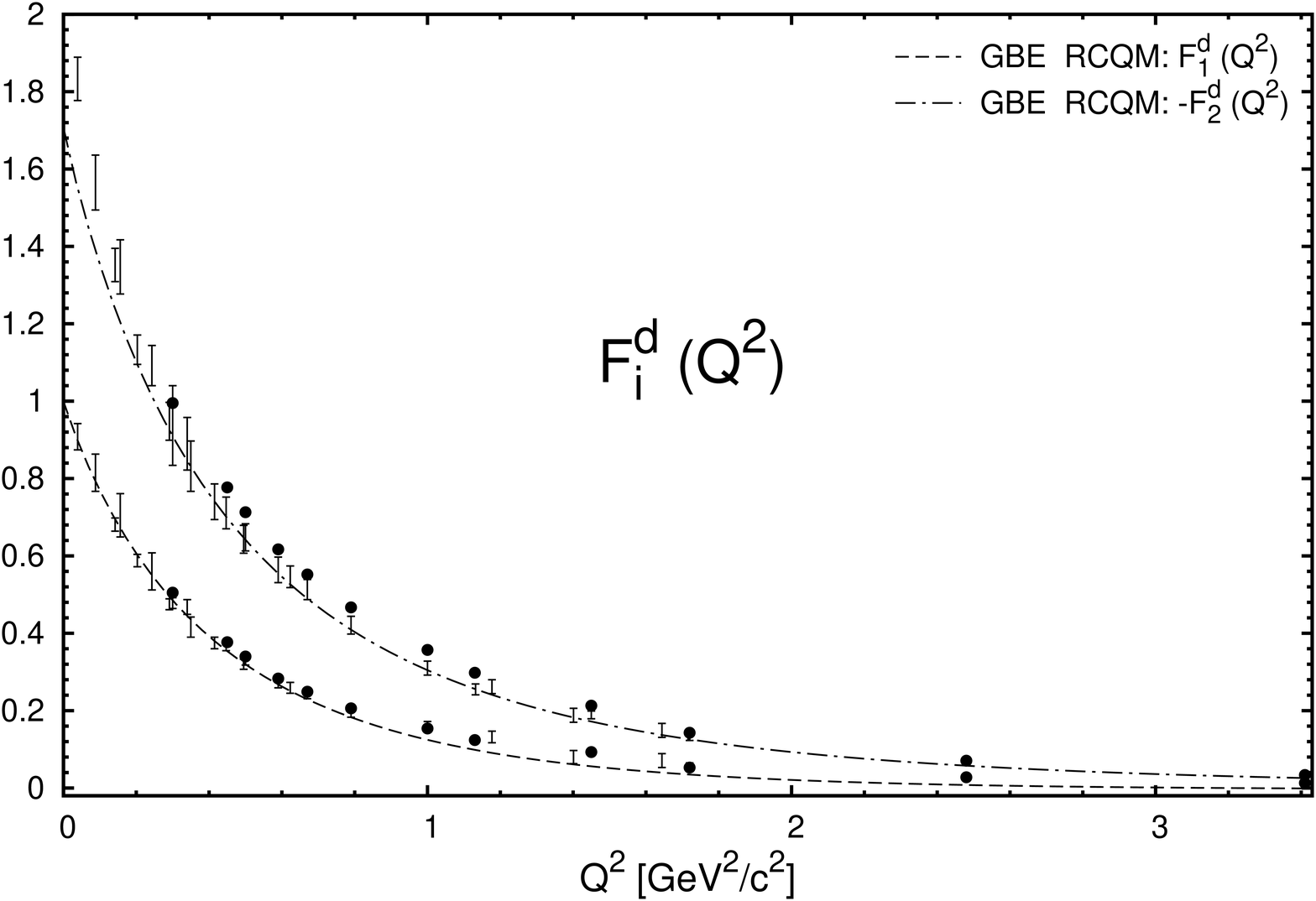}
\caption{Momentum dependences of the $u$- and $d$-flavor contributions to the nucleon
Dirac and Pauli form factors defined in Eq.~(\ref{fiq}) as predicted by the GBE RCQM
in comparison to phenomenological data from refs.~\cite{Cates:2011pz,Diehl:2013xca}; same
symbols as in Fig.~\ref{ffs}.}
\label{fi} 
\vspace{-3mm}
\end{figure*}

The direct predictions of the GBE RCQM for the flavor-separated Dirac and Pauli form factors
defined by
\begin{equation}
F_i^u=2F_i^p+F_i^n\,, \quad\,\, F_i^d=F_i^p+2F_i^n\,, \quad\,\, i=1,2
\label{fiq}
\end{equation}
are shown in Fig.~\ref{fi} in comparison to the phenomenological analyses of
refs.~\cite{Cates:2011pz,Diehl:2013xca}. Again a reasonable agreement is observed.

\begin{figure}[h]
\includegraphics[clip=,width=8.9cm]{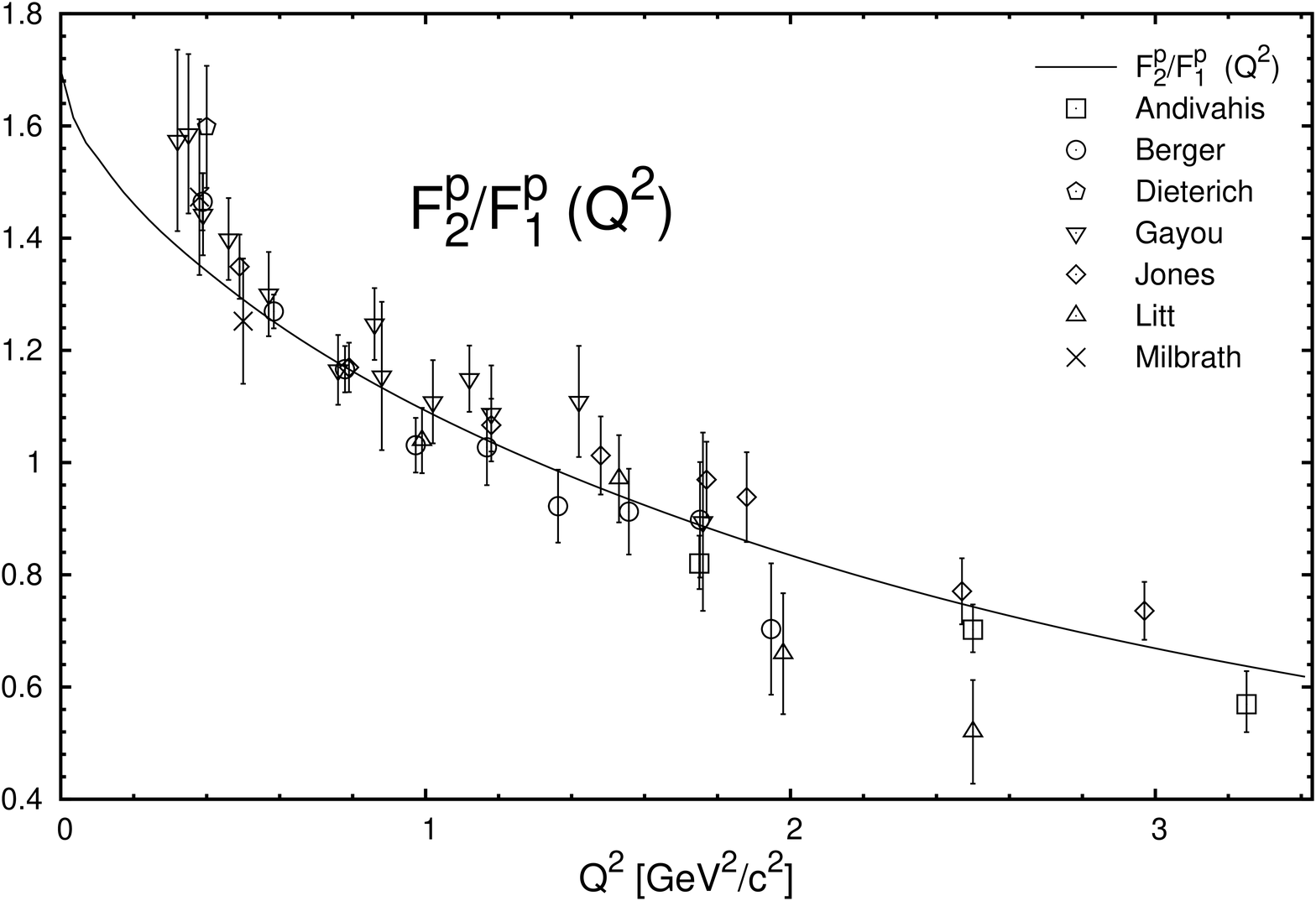} \\[-2mm]
\includegraphics[clip=,width=8.9cm]{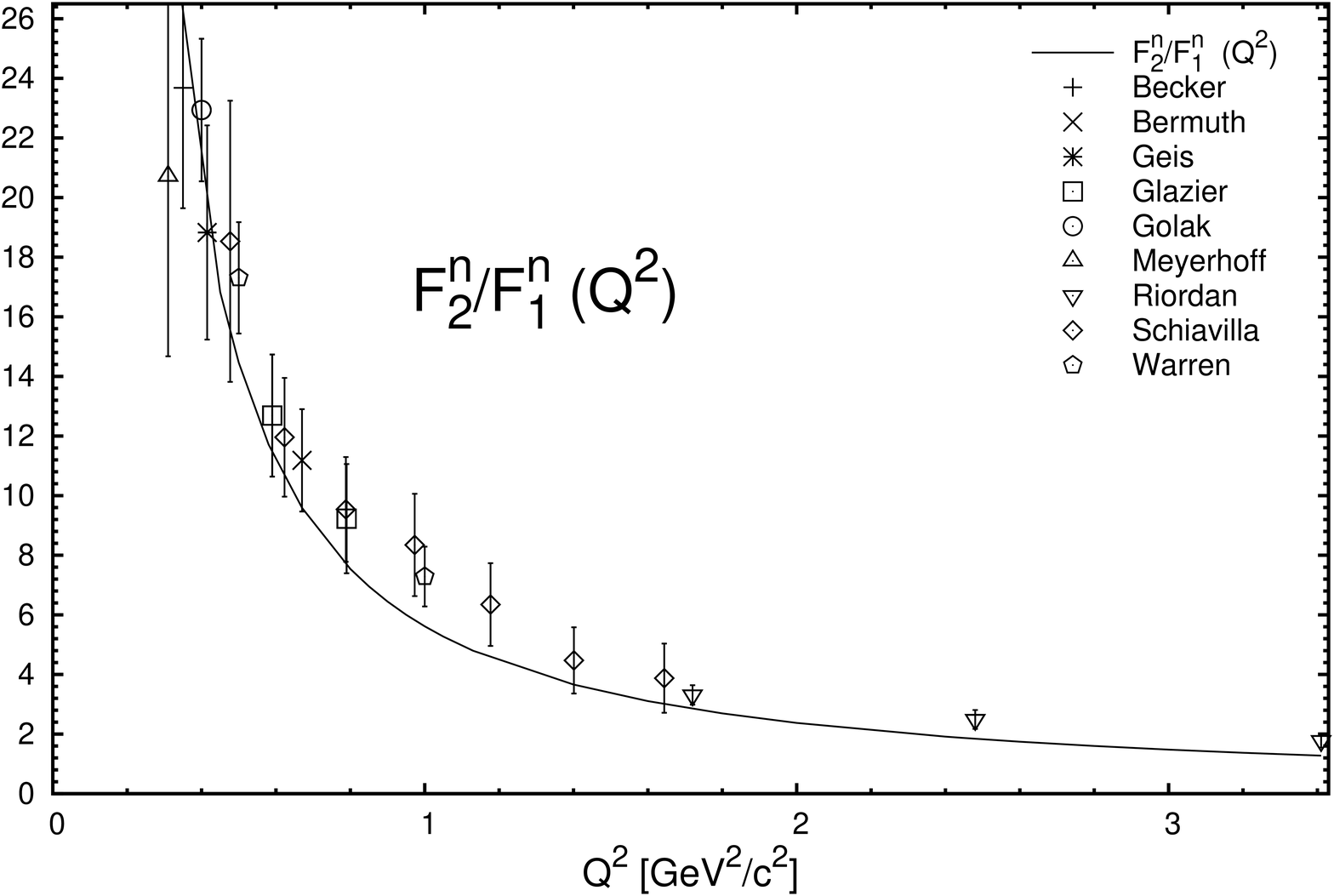}
\caption{Ratios of Pauli to Dirac form factors for the proton (top) and neutron
(bottom) as predicted by the GBE RCQM in comparison to experimental data as indicated.}
\label{PtoDffs} 
\vspace{-5mm}
\end{figure}

\begin{figure}[b]
\includegraphics[clip=,width=9cm]{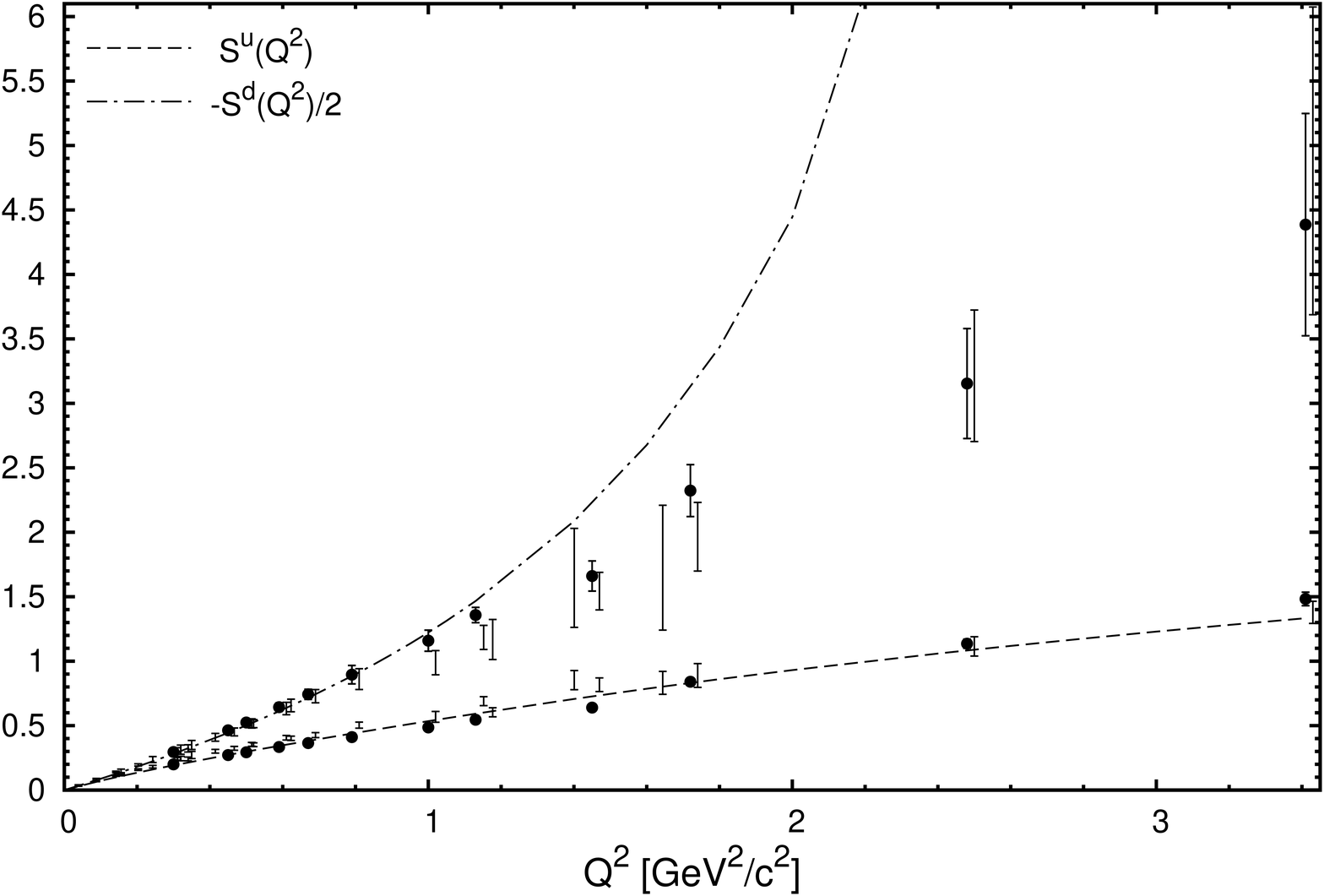}\\
\caption{The functions $S^q(Q^2)$ defined in Eq.~(\ref{Sq}), representing the ratios
of the $u$- and $d$-flavor contributions to the Pauli and Dirac form factors in
Fig.~\ref{fi}, as predicted by the GBE RCQM in comparison to phenomenological data
from refs.~\cite{Cates:2011pz,Diehl:2013xca}. The symbols are the same as in Fig.~\ref{ffs},
where the vertical bars corresponding to ref.~\cite{Diehl:2013xca} are slightly displaced
to the right in order to discriminate between the uncertainties in both data sets,
in case the $Q^2$ values on the abscissa coincide.}
\label{sq} 
\end{figure}

Often ratios of electromagnetic form factors are considered, since they are directly
accessible by experiment and provide a more discriminative insight into the $Q^2$
dependences with smaller uncertainties. In Fig.~\ref{PtoDffs} we first show the
predictions of the GBE RCQM for the
$F_2/F_1$ ratios for the proton and the neutron. The analogous ratios of the flavor-separated form factors are depicted in Fig.~\ref{sq} by means of the functions
\begin{equation}
S^q(Q^2)=Q^2 \frac{F_2^q(Q^2)}{F_1^q(Q^2)} \,, \quad\quad q=u,d \,.
\label{Sq}
\end{equation}

According to the phenomenological data both functions $S^q(Q^2)$ reflect an almost linear
rise with increasing $Q^2$ (i.e. constancy of the $Q^2$ dependences
of the ratios $F_2^q/F_1^q$). This behavior is met by the GBE RCQM only for the $u$-flavor.
The $d$-flavor ratio starts to depart from the phenomenological data at $Q^2\sim$1.5 GeV$^2$.

\begin{figure}[t]
\includegraphics[clip=,width=9cm]{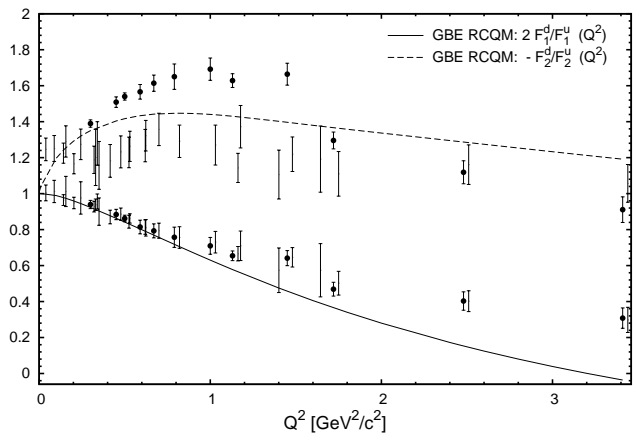}\\
\caption{Ratios of $u$- and $d$-flavor contributions to the Dirac and Pauli form factors
as predicted by the GBE RCQM in comparison to phenomenological data from
refs.~\cite{Cates:2011pz} and~\cite{Diehl:2013xca}. Same symbols and remarks as in
Fig.~\ref{sq}.}
\label{fdfu} 
\vspace{-2mm}
\end{figure}

Regarding the distinct $Q^2$ dependences of the flavor-separated Dirac and Pauli form
factors $F_i^u$ and $F_i^d$ we have also examined the ratios $F_i^d/F_i^u$ (see
Fig.~\ref{fdfu}). For both
$i=1,2$ they reasonably follow the phenomenological data. In the regime of low momentum
transfers this is especially true for $F_1^d/F_1^u$, where the data sets of
refs.~\cite{Cates:2011pz,Diehl:2013xca} also coincide. This is obviously not the case
for $F_2^d/F_2^u$. Our theoretical predictions just fall amidst these two phenomenological
data sets. We note that the recent analysis of ref.~\cite{Qattan:2012zf} has yielded
data of quite a similar behavior, and our theory perfectly agrees with them (see
Fig. 9 of ref.~\cite{Qattan:2012zf}). Anyway, here is a case where the phenomenological
analyses~\cite{Cates:2011pz,Qattan:2012zf,Diehl:2013xca} have produced different answers,
which do not seem to be compatible.

In any case, it seems to be well established from experiments that the $d$-contributions
$F_i^d$ fall off faster than the $u$-contributions $F_i^u$ towards higher momentum transfers,
a behavior that has sometimes been interpreted as an indication for diquark clustering
in the nucleons. However, our GBE RCQM, which is a genuine three-quark microscopic model
and has no reference to diquark configurations whatsoever (see its spatial probability distributions depicted in ref.~\cite{Melde:2008yr}), also produces these fall-offs in
overall agreement with the phenomenological data. The above reasoning regarding diquark
clustering thus appears questionable.

In view of the reasonable flavor contributions to the electromagnetic form factors found
for the GBE RCQM results, let us shortly summarize the main ingredients in our theory.
First of all,
we have very precise nucleon wave functions generated by an interacting mass operator
that contains a linear confinement, with a strength corresponding to the string tension
of QCD, and a hyperfine interaction that incorporates SB$\chi$S, leading to a specific
flavor dependence, contrary to a one-gluon-exchange hyperfine interaction (see the
corresponding discussion in ref.~\cite{Glozman:1998fs}).
The nucleon wave functions contain non-zero orbital angular momenta as well as
mixed-symmetric spatial wave-function parts; the latter are most important for reproducing
the neutron electromagnetic form factors, even though the non-symmetric
spatial components are relatively
small. Furthermore, due to the use of point-form relativistic quantum mechanics,
the matrix elements of the electromagnetic current operator are strictly
frame-independent~\cite{Melde:2004qu}.
We may calculate the electromagnetic form factors in the Breit frame, the laboratory frame,
or in any other frame, as we can perform the Lorentz transformations exactly in the
point form. Our specific construction of the electromagnetic current operator -- according
to the point-form spectator model -- also guarantees current
conservation~\cite{Melde:2007zz}.

The GBE RCQM for the nucleons relies on three-quark configurations only. The underlying
interaction Lagrangian is just built by coupling valence-quark fields with Goldstone
bosons. No explicit mesonic effects or more-quark components are introduced. From our
previous studies we have learned that relativistic (boost) effects are most important
in the reproduction of the nucleon electromagnetic form
factors~\cite{Wagenbrunn:2000es,Boffi:2001zb}. This is even true for the
quantities extracted at or near zero momentum transfers, i.e. the magnetic moments and
electric radii~\cite{Berger:2004yi}. For the success of the GBE RCQM we identify as
the essential symmetry ingredients the respect of SB$\chi$S of low-energy QCD and
Lorentz invariance.

\begin{acknowledgments}
This work was supported by the Austrian Science Fund, FWF, through the Doctoral
Program on {\it Hadrons in Vacuum, Nuclei, and Stars} (FWF DK W1203-N16). The authors are
grateful to B. Wojtsekhowski for providing them with an extensive experimental data base of
elastic nucleon electromagnetic form factors.
\end{acknowledgments}
\vspace{-3mm}
\addcontentsline{toc}{chapter}{Bibliography}

\end{document}